\documentclass[journal]{IEEEtran}

\usepackage{times}
\usepackage{epsfig}
\usepackage{graphicx}
\usepackage{amsmath}
\usepackage{amssymb}
\usepackage{subfigure}
\usepackage{gensymb}
\usepackage{algorithm}
\usepackage{algorithmic}
\usepackage{bm}

\DeclareFontFamily{OT1}{pzc}{}
\DeclareFontShape{OT1}{pzc}{m}{it}{<-> s * [1.10] pzcmi7t}{}
\DeclareMathAlphabet{\mathpzc}{OT1}{pzc}{m}{it}

\begin{document}

\title{Deep Amended Gradient Descent for Efficient Spectral Reconstruction from Single RGB Images}



\author{Zhiyu Zhu, Hui Liu, Junhui Hou, \textit{Senior Member, IEEE}, Sen Jia,  \textit{Senior Member, IEEE}, \\and Qingfu Zhang \textit{Fellow, IEEE} 
\thanks{This work was supported by the Hong Kong Research Grants Council under Grants 9048123 (CityU 21211518) and 9042820 (CityU 11219019). 
}
\thanks{Z. Zhu, H. Liu, J. Hou, and Q. Zhang are with the Department of Computer Science, City University of Hong Kong, Hong Kong. E-mail: \{zhiyuzhu2-c, hliu99-c\}@my.cityu.edu.hk, \{jh.hou, qingfu.zhang\}@cityu.edu.hk}
\thanks{S. Jia is with the School of Computer Science and
Software Engineering, Shenzhen University, China. E-mail: senjia@szu.edu.cn}}
\maketitle

\begin{abstract}
This paper investigates the problem of recovering hyperspectral (HS) images from single RGB images. To tackle such a severely ill-posed problem, 
we propose a physically-interpretable, compact, efficient, and end-to-end learning-based framework, namely AGD-Net. 
Precisely, by taking advantage of the imaging process, 
we first formulate the problem explicitly 
based on the classic gradient descent algorithm. 
Then, we design a lightweight neural network with a multi-stage architecture 
to mimic the formed amended gradient descent process, in which efficient convolution and novel spectral zero-mean normalization are proposed to effectively extract spatial-spectral features for regressing an initialization, a basic gradient, and an incremental gradient. 
Besides, 
based on the approximate low-rank property of HS images, we propose a novel rank loss 
to promote the similarity between the global structures of reconstructed  and ground-truth HS images, which is optimized with our singular value weighting strategy during training. 
Moreover, AGD-Net, a single network after one-time training, is flexible to handle the reconstruction with various spectral response functions.
Extensive experiments over three commonly-used benchmark datasets demonstrate that AGD-Net can improve the reconstruction quality by more than 1.0 dB on average while saving 67$\times$ parameters and 32$\times$ FLOPs, compared with state-of-the-art methods. The code will be publicly available at https://github.com/zbzhzhy/GD-Net. 

\end{abstract}


\IEEEpeerreviewmaketitle

\section{Introduction}
\IEEEPARstart{O}{wing}  to the dense sampling in the spectral domain, hyperspectral (HS) images 
can provide more accurate and faithful measurements towards the real-world scenes/objects than traditional RGB images. Such rich spectral information will be beneficial to 
various vision-based applications, such as tracking \cite{xiong2020material}, segmentation \cite{nalepa2019validating}, and detection \cite{mayer2003object,sharma2020hyperspectral}. 
However, the acquisition of HS images is costly, 
which severely limits the wide deployment of HS image-based applications. 

Instead of relying on the development of hardware,
many computational methods, such as compressive sensing-based \cite{wang2020dnu,wang2016adaptive,wang2018high,fu2018snapshot,zheng2021deep,he2020fast,meng2020end}
, HS and RGB image fusion \cite{qu2018unsupervised,dong2016hyperspectral,yao2020cross,liu2021hyperspectral,xu2019nonlocal}, \cite{wen2018deepcasd},
 single RGB image-based \cite{galliani2017learned,arad2020ntire,peng2020residual,xiong2017hscnn}, and
spatial super-resolution \cite{li2017hyperspectral,hu2017hyperspectral,liu2021spectral,li2021exploring},  \cite{zhang2012super},
have been proposed to acquire HS images in an affordable and convenient manner.  
Particularly, reconstructing HS images from single RGB images, which does not require specially-designed acquisition hardware, is a promising direction. 
Owing to the strong ability of learning representations, 
deep neural network (DNN)-based methods have recently been proposed to address this challenging task 
\cite{shi2018hscnn+,zhang2020pixel}. For example, Zhang \textit{et al.} \cite{zhang2020pixel} proposed pixel-aware deep learning framework for spectral upsampling. Li \textit{et al.} \cite{li2020adaptive,li2020hybrid} introduced the spectral and spatial attention mechanism into the reconstruction process. See Sec. \ref{sec:RW} for more details.
However, most of existing DNN-based spectral reconstruction methods adopt architectures for general purposes, 
and neglect the unique characteristics of this task, e.g., the specific relationship between HS and RGB images, which may compromise their performance. Second, the majority of them trained with RGB images acquired via a typical spectral response function (SRF) cannot handle RGB images via a different SRF during inference, which limits their use in practice to some extent. 
In addition, existing DNN-based methods were usually trained with pixel-wise loss functions, 
which fail to capture the global structure of HS images, i.e., the relationship among spectral bands.


In this paper, we propose a novel DNN-based framework, which is highlighted with compact, efficient, interpretable, and effective characteristics, for the reconstruction of HS images from single RGB images in an end-to-end fashion.
Specifically, based on the specific relationship between RGB and HS images, we first explicitly formulate the problem as amended gradient descent (AGD) progress, 
which boils down to determining an initialization, a basic gradient, and an incremental gradient. 
Then, we propose 
AGD-Net with a multi-stage structure to mimic the AGD process, in which with the initialization learned, the basic and incremental gradients are adaptively and progressively learned at each stage by embedding the spatial-spectral information of input RGB images via memory- and computationally-efficient convolution and novel spectral zero-mean normalization. 
To exploit the global structure of HS images, we also propose a novel rank loss, which is optimized via a singular value weighting strategy during training.
Thanks to the interpretable architecture, we extend AGD-Net to enable a single network after one-time training can 
handle input RGB images generated with different SRFs.
Extensive experimental results demonstrate the significant superiority of AGD-Net over state-of-the-art methods, i.e., AGD-Net reconstructs HS images with much higher quality but at lower memory and computational costs.

The rest of this paper is organized as follows. Sec. \ref{sec:RW} briefly reviews existing methods for HS image reconstruction. Sec. \ref{sec:formulation} formulates the problem. Sec. \ref{sec:method} presents the proposed framework, followed by extensive experimental results as well as analyses in Sec. \ref{sec:EXP}. Finally, Sec. \ref{sec:CON} concludes this paper.




  
\section{Related Work}
\label{sec:RW}

 In the following, we briefly review the existing works on the reconstruction of HS images from single RGB images.
 \subsection{Traditional Methods}
Many traditional methods assume that HS images lie in a low-dimensional subspace and explore the map between RGB images and subspace coordinates. For example, Nguyen \textit{et al.} \cite{nguyen2014training} leveraged RGB white-balancing
to normalize the scene illumination to recover the scene reflectance. Arad \textit{et al.} \cite{arad2016sparse} proposed a sparse coding-based method, which learns an over-complete dictionary of HS images to describe the novel RGB images. Then Aeschbacher \textit{et al.} \cite{aeschbacher2017defense} further improved it  
 through a shallow A+-based method \cite{timofte2014a+}. Jia \textit{et al.} \cite{jia2017rgb} exploited the 3D embedded space where the natural scene spectra reside and learned an accurate non-linear mapping from RGB images to 3D embeddings.
Heikkinen \textit{et al.} \cite{heikkinen2018spectral} estimated the spectral subspace coordinates via a scalar-valued Gaussian process regression with an-isotropic or combination kernels.  Gao \textit{et al.} \cite{gao2020spectral} proposed a joint sparse and low-rank dictionary learning method for the reconstruction of HS images from single RGB images.

\subsection{DNN-based Methods}
 On the basis of the impressive representation ability of DNNs, many DNN-based methods have been proposed to reconstruct HS images from single RGB images. For example,  Xiong \textit{et al.} \cite{xiong2017hscnn} proposed a DNN-based method, namely HSCNN, for the reconstruction of HS images from RGB images or measurements obtained via compressive sensing, which mainly aims to enhance the spectral signatures constructed by 
 a simple interpolation or CS reconstruction. 
  Shi \textit{et al.} \cite{shi2018hscnn+} further improved HSCNN by replacing all predefined upsampling operators with residual blocks and introduced dense connections with a cross-scale fusion scheme to facilitate the feature extraction process. Gewali \textit{et al.} \cite{gewali2019spectral} utilized DNNs to optimize multispectral bands and hyperspectral recovery simultaneously to achieve more accurate HS image reconstruction. Fu \textit{et al.} \cite{fu2018spectral} modeled HS image reconstruction by exploring non-negative structured information and utilized multiple spare dictionary to learn a more compact basis representation. Berk \textit{et al.} \cite{kaya2019towards} trained multiple models to reconstruct HS images from RGB images captured with different SRFs, and they also trained an additional model to select different models during real-world applications. 
 Li \textit{et al.} \cite{li2020adaptive} also proposed an attention-based method which utilizes both channel attention and spatial non-local attention. 
 Based on the assumption that pixels in an HS image belong to different categories or spatial positions and often require distinct mapping functions, Zhang \textit{et al.} \cite{zhang2020pixel}  proposed a pixel-aware deep function-mixture network, which learns different bias functions and then linearly mixes them up according to pixel-level weights.
 Aitor \textit{et al.} \cite{alvarez2017adversarial} treated HS image reconstruction as an image to image mapping problem and applied a generative adversarial network to capture spatial semantics. Yan \textit{et al.} \cite{yan2020reconstruction} introduced prior category information to generate distinct spectral data of objects via a U-Net-based architecture. Zhao \textit{et al.} \cite{zhao2020hierarchical} presented a hierarchical regression network with a pixel shuffle layer. Fu \textit{et al.} \cite{fu2018joint} developed an SRF selection layer to retrieve the optimal response function for HS image reconstruction.  Peng \textit{et al.} \cite{peng2020residual} introduced a pixel-wise attention module for boosting reconstruction performance. Galliani \textit{et al.} \cite{galliani2017learned} utilized a densely connected U-Net-based architecture for HS images reconstruction. However, the performance of the above-mentioned methods is still limited, due to insufficient modeling towards the problem. Besides, although these 
 methods attempt to build reconstruction processes with physical meaning, the adopted 
 architectures for general purposes seriously restrict their interpretability. 
\subsection{Algorithm Unrolling-based Methods}
As our deep learning-based framework is driven by model-based optimization, we also briefly review some related works under this stream. 
Since Gregor and LeCun \cite{gregor2010learning} developed a sparse coding-based algorithm unrolling technique, a number of unrolling iterative algorithms with DNNs have been proposed for various image reconstruction, 
such as single RGB image super-resolution \cite{zhang2019deep,deng2019deep}, compressive sensing \cite{sun2016deep,ma2019deep}, and image fusion \cite{xie2019multispectral}. Generally, this kind of methods solves inverse problems via unfolding optimization steps and applying DNNs to solve them in a data-driven manner. The main differences among those methods lie in the formulation of an inverse problem as well as adopted optimization algorithms, which will result in various network architectures. 
For example, Lohit \textit{et al.} \cite{lohit2019unrolled} unrolled a projected gradient descent algorithm for HS image pan-sharpening. Wen \textit{et al.} \cite{wen2018deepcasd} utilized a deep coupled analysis and synthesis dictionary-based network for HS image super-resolution. Wang \textit{et al.} \cite{wang2020dnu,wang2019hyperspectral} unfolded a half quadratic splitting algorithm using DNNs for coded aperture snapshot spectral imaging. We refer readers to \cite{monga2021algorithm} for the comprehensive survey on algorithm unrolling.

\begin{figure*}
	\centering
	\includegraphics[width=0.99\linewidth, trim=0 0 0 0, clip]{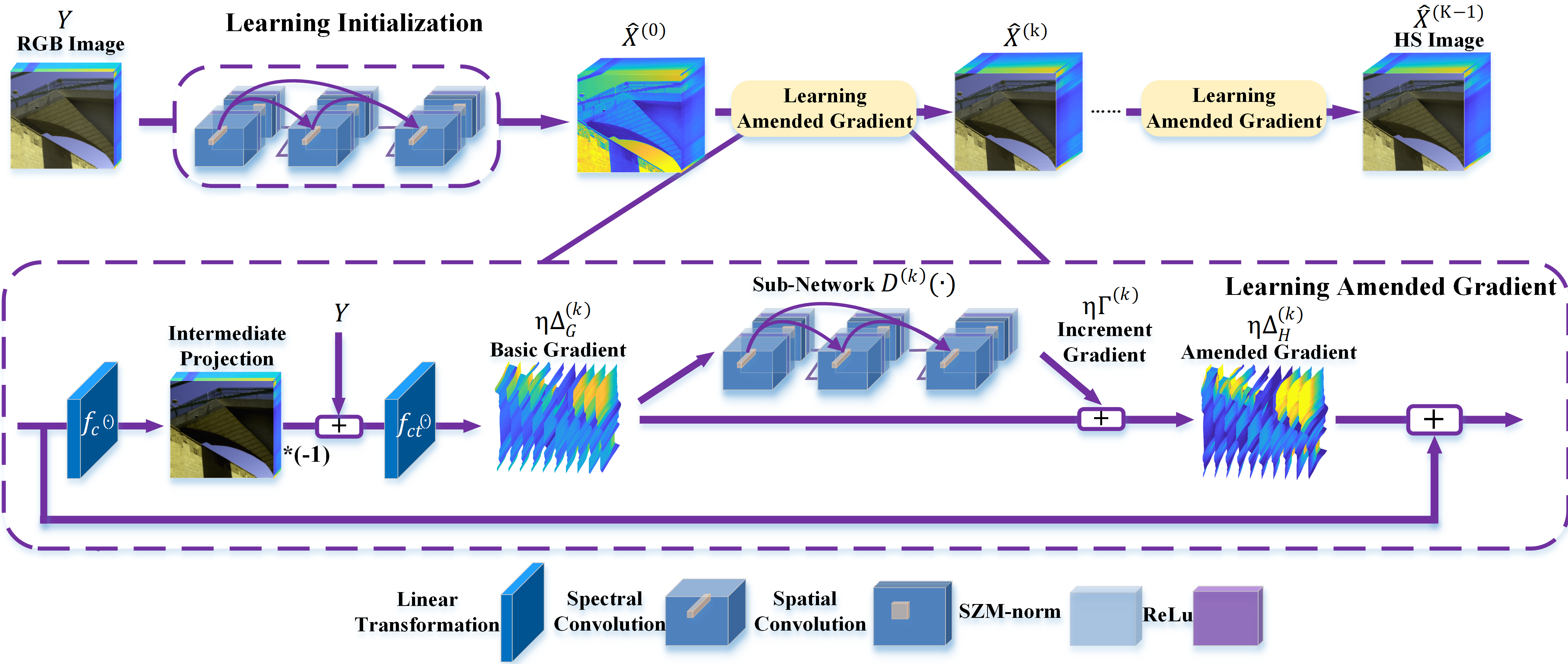}
	\caption{Illustration of the flowchart of the proposed AGD-Net, a compact, interpretable, and end-to-end neural network with a multi-stage architecture, for the reconstruction of HS images from single RGB images. AGD-Net mimics an amended gradient descent process to solve the formed ill-posed inverse problem.  Each stage consists of two modules, namely learning initialization and learning amended gradient. 
	}
 \label{fig:flowchart}
\end{figure*}

\section{Problem Formulation}
\label{sec:formulation}

Denote by $\mathbf{Y}\in\mathbb{R}^{3\times hw}$ the vectorial representation of an RGB image of spatial dimensions $h\times w$, and by $\mathbf{X} \in \mathbb{R}^{s\times hw}$ the corresponding HS image with $s$  ($s\gg3$) spectral bands  to be reconstructed.  
The relationship between  $\mathbf{X}$ and $\mathbf{Y}$ 
can be generally formulated as
\begin{align}
\label{eq:observ}
	\mathbf{Y} = \mathbf{C}\mathbf{X} + \mathbf{N}_y,
\end{align}
where 
$\mathbf{C} \in \mathbb{R}^{3 \times s}$ is the spectral response function (SRF), 
and $\mathbf{N}_y \in \mathbb{R}^{3\times hw} $ is the noise. 
Simply, under the assumption that the noise is normally distributed, we can recover $\mathbf{X}$ 
from $\mathbf{Y}$ 
by optimizing the following 
problem formulated from Eq. (\ref{eq:observ}):
\begin{align}
\label{eq:target2}
\min_\mathbf{\mathbf{X}} \quad \mathcal{G}\left(\mathbf{X}\right) = \frac{1}{2} \|\mathbf{Y}-\mathbf{C}\mathbf{X}\|_F^2,
\end{align}
where $\|\cdot\|_F$ is Frobenius norm of a matrix. 
Moreover, with an initial guess $\mathbf{X}^{(0)}$, we can solve Eq. (\ref{eq:target2}) with the classic gradient descent (GD) algorithm, and at the $k$-th ($1 \leq k\leq K-1$) step, we have 
\begin{align}
\label{eq:target6}
\mathbf{X}^{(k)} = \mathbf{X}^{(k-1)} +  \eta  \Delta_\mathcal{G}\left(\mathbf{X}^{(k-1)}\right),
\end{align}
where $\eta$ is the step size 
and $\Delta_\mathcal{G}(\cdot)$ is the operator of computing the derivative 
of $\mathcal{G}(\cdot)$, i.e., 
\begin{align}
\label{eq:target5}
\Delta_\mathcal{G}\left(\mathbf{X}^{(k)}\right) = \mathbf{C}^\textsf{T}\left(\mathbf{Y} - \mathbf{C}\mathbf{X}^{(k)}\right).
\end{align}

Unfortunately, it is almost impossible to obtain a feasible solution by means of such a simple optimization process, due to the severely ill-posed behavior of the problem in Eq. (\ref{eq:target2}), i.e., there are numerous trivial solutions.  In addition, the performance of such a scheme highly depends on the initialization.  From the perspective of gradient space, the reason could be interpreted as that the gradient cannot decrease either along with an optimal path or from an appropriate starting point to the global minimum or a good local minimum during the iteration process.
Therefore, to make the gradient process effective, an intuitive thought is that 
we can find an appropriate initialization and amend the gradient at each step of the iteration process 
That is,  instead of Eq. (\ref{eq:target5}), we can generally express the gradient at the $k$-th step as 
\begin{equation}
 \label{eq:target8}
 \Delta_\mathcal{H}\left(\mathbf{X}^{(k)}\right) =\Delta_\mathcal{G}\left(\mathbf{X}^{(k)}\right) + {\bf\Gamma}^{(k)}, 
 \end{equation}
where $\Delta_\mathcal{H}\left(\mathbf{X}^{(k)}\right)$ is the amended gradient, and ${\bf\Gamma}^{(k)}\in\mathbb{R}^{s\times hw}$ is the incremental gradient. Accordingly, we obtain the amended gradient descent process as
\begin{equation}
    \label{eq:GDprocess}
    \mathbf{X}^{(k)} = \mathbf{X}^{(k-1)} +  \eta  \Delta_\mathcal{H}\left(\mathbf{X}^{(k-1)}\right).
\end{equation}

\section{Proposed Method}
 \label{sec:method}

 

Motivated by the intuitive and explicit formulation in Sec. \ref{sec:formulation}, as illustrated in Fig. \ref{fig:flowchart}, we propose a novel end-to-end and lightweight DNN-based framework, namely AGD-Net, which mimics the amended gradient descent process, to achieve the reconstruction of HS images from single RGB images. To be specific, with the initialization $\mathbf{X}^{(0)}$ learned, 
we progressively learn the basic gradient $\Delta_{\mathcal{G}}(\mathbf{X}^{(k)})$ and the incremental gradient ${\bf\Gamma}^{(k)}$ via a multi-stage architecture, in which the spatial-spectral information of the input RGB image is effectively and efficiently embedded. 
Besides, we propose a global structure-aware loss function to train AGD-Net end-to-end.  In what follows, we detail each module. 
 
\subsection{Learning Initialization}
\label{subsec:initial}
This module aims to learn an appropriate initialization $\mathbf{X}^{(0)}$ as the starting point of the gradient descent process. 
We adopt a 
densely-connected convolutional neural network (CNN) to extract spatial-spectral information of $\mathbf{Y}$ to regress $\mathbf{X}^{(0)}$.
Specifically, to learn feature representations efficiently and effectively, 
we adopt a series of  memory- and computational-efficient spectral-spatial separable convolution \cite{ zhu2020hyperspectral}, 
which applies
two kinds of sequentially connected convolution, namely 1D spectral convolution and 2D spatial convolution, with an in-between activation function.
Specifically, the former applies kernels of size 1$\times$1 in 1D spectral/channel space for embedding spectral information, while the latter applies kernels of size $3\times 3$ in the 2D spatial space for embedding spatial information.
Moreover, to emphasize high-frequency spectral information and regularize the intermediate feature away from overfitting, we propose spectral zero-mean normalization (SZM-norm), which enforces the vector formed by the features from different channels but at an identical spatial position to have a zero-mean, i.e., 
 \begin{align}
 \mathcal{Z}\left(x_{j,m,n}\right) = x_{j,m,n} - \frac{1}{s}\sum_{j=1}^{s}\left(x_{j,m,n}\right),
 \end{align}
 where $\mathcal{Z}(\cdot)$ denotes SZM-norm, $x_{j,m,n}$ is the $(m,n)$-th element of the feature map of the  
 $j$-th  channel.  We will experimentally validate the effectiveness of this initialization module and the SZM-norm in the following Table \ref{tab:ablation2}.
 

\subsection{Learning the Amended Gradient}
In this module, we aim to learn an amended gradient, which is the sum of a basic gradient and an incremental gradient. 
 
 	
\subsubsection{Basic gradient} As formulated in Eq. (\ref{eq:target5}), 
 the SRF $\mathbf{C}$ and its transpose $\mathbf{C^\textsf{T}}$ in Eq. (\ref{eq:target5}) actually act as the linear projection in pixel-wise, 
 and we thus 
 simulate $\mathbf{C}$ with a convolutional layer denoted as $f_c\left(\cdot, \bm{\theta}_c^{(k)}\right)$, and $\mathbf{C}^{\textsf{T}}$ with a corresponding deconvolutional layer denoted as $f_{ct}\left(\cdot, \bm{\theta}_{ct}^{(k)}\right)$ for the back projection,   
 where $\bm{\theta}_{c}^{(k)}$ and $\bm{\theta}_{ct}^{(k)}$ are the sets of parameters to be learned. 
Accordingly, the scaled basic gradient\footnote{The scaled basic gradient refers to the product of the step size $\eta$ and $\Delta_{\mathcal{G}(\cdot)}$} is derived as 
\begin{align}
 	\label{eq:target10}
\eta\Delta_\mathcal{G}\left(\widehat{\mathbf{X}}^{(k)}\right) =  f_{ct}\left(\mathbf{Y} -  f_c\left(\widehat{\mathbf{X}}^{(k)}, \bm{\theta}_{c}^{(k)}\right), \bm{\theta}_{ct}^{(k)}\right),
\end{align}
where $\widehat{\mathbf{X}}^{(k)}$ is the intermediate HS image reconstructed at the $k$-th stage.
Note that these two convolutional layers are not followed by an activation function in order to preserve the linear property of these transformations. 

In addition, considering that the linear projection layers $f_c\left(\cdot, \bm{\theta}_c^{(k)}\right)$ in all stages have the same purpose, 
i.e., adaptively learning the SRF, and we only explicitly supervise $f_c\left(\cdot, \bm{\theta}_c^{(K-1)}\right)$ during training,  we apply shared parameters to these layers, i.e., $\bm{\theta}_c^{(1)}=\bm{\theta}_c^{(2)}=\cdots=\bm{\theta}_c^{(K-1)}$, 
to guarantee the error $\mathbf{E}^{(k)}=\left( \mathbf{Y}-f_c\left(\widehat{\mathbf{X}}^{(k)}, \bm{\theta}_c^{(k)}\right)\right)$ can be correctly calculated at all stages.  
We experimentally validate the effectiveness of such a weight sharing strategy in Table \ref{tab:ablation2}. 
 	 	
\subsubsection{Incremental gradient} 
Considering that both the basic gradient and the incremental gradient are distributed in gradient space, we directly learn the incremental gradient ${\bf\Gamma}^{(k)}$ from $\eta\Delta_{\mathcal{G}}(\widehat{\mathbf{X}}^{(k)})$ by using a sub-network denoted as $\mathcal{D}^{(k)}\left(\cdot, \bm{\theta}^{(k)}\right)$, i.e., 
 \begin{equation}
     \eta{\bf\Gamma}^{(k)}=\mathcal{D}^{(k)}\left(\eta\Delta_{\mathcal{G}}(\widehat{\mathbf{X}}^{(k)}), \bm{\theta}^{(k)}\right),
 \end{equation}
where $\bm{\theta}^{(k)}$ is the set of parameters at the $k$-th stage to be learned.
For simplicity, we adopt the same network architecture as that in Sec. \ref{subsec:initial} but different parameters to realize $\mathcal{D}^{(k)}(\cdot, \cdot)$, whose architecture details are summarized in Table \ref{tab:Keyfeature}.
 
According to Eq. (\ref{eq:target8}), we can derive the amended gradient at the $k$-th stage as 
 	\begin{align}
 	\label{eq:target9}
 \eta \Delta_\mathcal{H}\left(\widehat{\mathbf{X}}^{(k)}\right) &= \\ &\eta\Delta_\mathcal{G}\left(\widehat{\mathbf{X}}^{(k)}\right) +\mathcal{D}^{(k)}\left(\eta\Delta_\mathcal{G}(\widehat{\mathbf{X}}^{(k)}),\bm{\theta}^{(k)}\right).\nonumber
 	\end{align}
It can be seen that Eq. (\ref{eq:target9}) has the same form as residual learning \cite{he2016deep}, and thus the advantages of residual learning will be inherited. 
Note that we remove all the bias of the convolutional layers 
in $\mathcal{D}^{(k)}\left(\cdot,\cdot\right)$. The reason is that the error $\mathbf{E}^{(k)}$ 
also measures the differences between reconstructed and ground-truth HS images, and when it reaches zero, the optimization process has found an appropriate reconstructed HS image with respect to Eq. (\ref{eq:observ}). Then, 
the updating of the HS image should be terminated, requiring the amended gradient be zero, which is equivalent to that the sub-network must pass through  origin:
 	\begin{align}
 	\label{eq:target11}
 	\mathcal{D}^{(k)}\left(\mathbf{0} ,\bm{\theta}^{(k)}\right) \equiv \mathbf{0},
 	\end{align}
where $\mathbf{0}$ is a matrix with all elements equal to zero.

 \begin{table*}
 	\caption{\label{tab:Keyfeature} The architecture details of  $\mathcal{D}^{(k)}\left(\cdot,\theta^{(k)}\right)$. 
 	}
 	\centering
\resizebox{\textwidth}{!}{
 	\begin{tabular}{l c c c c c c}
 		\hline	\hline
 			&Kernel shape 	  						& \# Input Channels 	&\# Output Channels &Output shape               & ReLU & SZM-norm \\
 		\hline
 		\textbf{The $l$-th Spectral-spatial separable convolutional layer  $l\in [1,4]$}\\
 		\quad Spectral convolution										&62$l$ $\times$62$\times$1$\times$1			&62$ l$			&62					&128$\times$128$\times$62	&$\checkmark$&$\checkmark$\\
 		\quad Spatial convolution										&62$\times$1$\times$3$\times$3			&62						&62 				&128$\times$128$\times$62	&$\checkmark$&$\checkmark$\\
 		\textbf{Spectral-spatial separable convolution (without activation)}\\
 		\quad Spectral convolution& 310$\times$31$\times$1$\times$1	&310					&31					&128$\times$128$\times$31	&--&$\checkmark$\\
 		\quad Spatial convolution&31$\times$1$\times$3$\times$3	&31						&31 					&128$\times$128$\times$31	&--&$\checkmark$\\
 		\hline	\hline
 	\end{tabular}}
 \end{table*}

\subsection{Global Structure-aware Loss Function} 
To train the AGD-Net, basically, we adopt the following pixel-wise loss function, i.e., 
\begin{align}
 \label{eq:Objective1}
  &\mathcal{L}_P\left(\mathbf{\widehat{X}},\mathbf{X}\right) =\mathcal{L}_F\left(\mathbf{\widehat{X}},\mathbf{X}\right) + \alpha \mathcal{L}_O\left(\mathbf{\widehat{X}},\mathbf{X}\right)  \\
  &= \frac{1}{3\times hw}\left\|f_c\left(\mathbf{\widehat{X}}\right)-\mathbf{Y}\right\|_F^2 +  \alpha \frac{1}{s\times hw}\left\|\mathbf{\widehat{X}}-\mathbf{X}\right\|_{1}, \nonumber%
 \end{align}
 where 
 $\|\cdot\|_1$ is the $\ell_1$ norm of a matrix, which computes the sum of the absolute values of all elements of a matrix, $\mathbf{\widehat{X}}$ and $\mathbf{X}$ are the reconstructed and ground-truth HS images, respectively,   $\alpha$ is the penalty parameter, 
 which is empirically set to 1, and $f_c(\cdot)$ is the convolutional layer projecting an HS image to the RGB image space. 
 Many previous works have experimentally demonstrated that the formed matrix from an HS image is an approximate low-rank matrix \cite{zhang2013hyperspectral,dian2019learning,dian2019hyperspectral,chang2020weighted,mei2018simultaneous,liu2021global}, i.e., the strong correlation among spectral bands.
 However, such a global structure of HS images cannot be captured by
 the pixel-wise loss in Eq. (\ref{eq:Objective1}).
To this end, we propose a rank loss $\mathcal{L}_R\left(\mathbf{\widehat{X}},\mathbf{X}\right)$. Specifically, we adopt a singular value weighting strategy to enforce the singular values of reconstructed HS images  to be exactly the same as those of the ground-truth HS images in a certain range $[\delta_l , \delta_h]$, based on the following two considerations:
\begin{itemize}
  \item relatively larger singular values correspond to more principal components (or low-frequency components of an image). However, for image reconstruction, the challenging issue lies in the recovery of high-frequency components, e.g., sharp details. Thus, we set an upper bound $\delta_h$ to promote the ability of the network in the learning of those details; and 
 
  \item the accuracy of eigenvectors corresponding to relatively small eigenvalues decreases. Thus, we set a lower bound $\delta_l$ to avoid utilizing the inaccurate eigenvectors. 
\end{itemize}
Algorithms \ref{alg:forward} and \ref{alg:backward} provide the forward and backward propagation of optimizing the rank loss during training, respectively.  

The overall loss function for training AGD-Net is finally written as
 \begin{equation}
 \label{eq:Objective}
 \mathcal{L}\left(\mathbf{\widehat{X}},\mathbf{X}\right) = \mathcal{L}_P\left(\mathbf{\widehat{X}},\mathbf{X}\right) + \beta \mathcal{L}_R\left(\mathbf{\widehat{X}},\mathbf{X}\right),
 \end{equation}
where the parameter $\beta$ is set to 1 to balance the two terms.

 \begin{algorithm}
 \caption{\label{alg:forward} Forward Propagation 
 }
 \begin{algorithmic}[1]
 \REQUIRE
 $\widehat{\mathbf{X}}$, 
  $\mathbf{X}  \in \mathbb{R}^{s\times hw}$, $h_1 (< h)$, $ w_1 (<w)$, $\delta_l$, and $\delta_h$  ($\delta_l<\delta_h$)\\
 \ENSURE 
 $\mathcal{L}_R\left(\mathbf{\widehat{X}},\mathbf{X}\right)$\\
   \STATE Partition $\mathbf{X}$ and $\widehat{\mathbf{X}}$ into $p$  patches of spatial dimensions $h_1\times w_1$, denoted as $\mathbf{X}_i \in \mathbb{R}^{s\times h_1w_1}$ and $\widehat{\mathbf{X}}_i \in \mathbb{R}^{s\times h_1w_1}$ ($1\leq i\leq p$), respectively. 
\FOR{$i\leftarrow 1:~ p$}
 \STATE		$[\mathbf{U}_i , \mathbf{\Lambda}_i ,\mathbf{V}_i]\leftarrow \textsf{SVD}\left(\mathbf{X}_i\right)$ and  $\left[\widehat{\mathbf{U}}_i, \widehat{\mathbf{\Lambda}}_i, \widehat{\mathbf{V}}_i\right]\leftarrow \textsf{SVD}(\mathbf{\widehat{X}}_i)$, 
 where $\mathbf{U}_i, \mathbf{\widehat{U}}_i\in\mathbb{R}^{s\times s}$, $\mathbf{\Lambda}_i, \mathbf{\widehat{\Lambda}}_i\in\mathbb{R}^{s\times s}$,  $\mathbf{V}_i, \mathbf{\widehat{V}}_i\in \mathbb{R}^{s\times h_1w_1}$, and $\textsf{SVD}(\cdot)$ performs the singular value decomposition \cite{golub2013matrix}.  Note  $\mathbf{\Lambda}_i$, $\widehat{\mathbf{U}}_i$, $\widehat{\mathbf{\Lambda}}_i$, and $\widehat{\mathbf{V}}_i$ will be saved for the reuse in the backward propagation in Algorithm \ref{alg:backward}.
 \STATE Initialize $\mathbf{Q}_i = \mathbf{0}\in\mathbb{R}^{s\times s}$.
 \FOR{$m\leftarrow 1:~s$}
 \IF{$\widehat{\lambda}_i^m>\delta_l$ \& $\widehat{\lambda}_i^m<\delta_h$}
  \STATE  $q_i^m\leftarrow 1$, where $\widehat{\lambda}_i^m$ is the $m$-th diagonal entry of $\widehat{\mathbf{\Lambda}}_i$, and $q_i^m$ is the $m$-th diagonal entry of 
  $\mathbf{Q}_i$. Note $\mathbf{Q}_i$ will be saved for the reuse in the backward propagation.
 \ELSE
 \STATE $q_i^m\leftarrow 0$.
 \ENDIF
 \ENDFOR
 \ENDFOR		
 		
 \STATE		$\mathcal{L}_R(\mathbf{\widehat{X}},\mathbf{X}) = \frac{1}{p\times s \times h_1 \times w_1} \sum_{i=1}^{p}\sum_{m=1}^{s} q_i^m $. 
 \end{algorithmic}
 \end{algorithm}
 \begin{algorithm}[h]
 \begin{algorithmic}[1]
 \caption{\label{alg:backward} Backward Propagation 
 }
 		\REQUIRE $\{\mathbf{\Lambda}_i\}_{i=1}^p$, $\{\widehat{\mathbf{U}}_i\}_{i=1}^p$, $\{\widehat{\mathbf{\Lambda}}_i\}_{i=1}^p$, $\{\widehat{\mathbf{V}}_i\}_{i=1}^p$, and $\{\mathbf{Q}_i\}_{i=1}^p$ \\
 		\ENSURE Gradient $\mathcal{G}(\widehat{\mathbf{X}}_i)$\\
\FOR{$i\leftarrow 1:~ p$}
 	\STATE	$\Delta_{\mathbf{\widehat{\Lambda}}_i} \leftarrow \frac{1}{s \times h_1 \times w_1} \left(\mathbf{\Lambda}_i-\mathbf{\widehat{\Lambda}}_i\right) \odot \mathbf{\widehat{\Lambda}}_i^{-1}\odot \mathbf{Q}_i$, where 
 	$\odot$ is the Hadamard product operator, and $\mathbf{\widehat{\Lambda}}_i^{-1}$ is the inverse matrix of $\mathbf{\widehat{\Lambda}}_i$. \\
 	\STATE	$\mathcal{G}(\widehat{\mathbf{X}}_i) \leftarrow \widehat{\mathbf{U}}_i  \Delta_{\mathbf{\widehat{\Lambda}}_i}  \widehat{\mathbf{V}}_i $\\
 \ENDFOR
 \end{algorithmic}
 \end{algorithm}

 \subsection{Flexible AGD-Net 
 }

In this section, 
we further extend AGD-Net for increasing its practicality and propose flexible AGD-Net (FAGD-Net), which is a single network that can handle data captured with various SRFs after only one-time training. Such an extension is enabled thanks to the interpretable architecture of AGD-Net. 

Specifically, to adapt various SRFs, we replace the learnable parameters of the linear projection layers in AGD-Net, i.e., $\bm{\theta}_{c}^{(k)}$ involved $f_c(\cdot,\cdot)$~$\forall k$, with explicit SRFs specified by the data, while keeping the remaining settings unchanged. We train FAGD-Net with RGB images acquired with various SRFs to augment its generalization ability. We carry out experiments to validate the effectiveness of FAGD-Net in Sec. \ref{sec:expextension}.

\begin{table*}
	\centering
	\caption{\label{tab:results1} Quantitative comparisons of different methods on the \textbf{HARVARD} dataset. ``$\uparrow$ (resp. $\downarrow$)" indicates that the larger (resp. smaller), the better.  For \# Params and \# FLOPs, the smaller, the more compact and efficient. The best results are highlighted in bold.}
\resizebox{0.7\textwidth}{!}{
	\begin{tabular}{l|cc|cccc}
		\hline
		\hline
		\textbf{Methods} & \# Params& \# FLOPs & PSNR $\uparrow$& ASSIM $\uparrow$& SAM $\downarrow$ & RMSE $\downarrow$ \\
		\hline
		\hline
		BI		 &--		&--	    & 23.71 &0.6945 &42.54   &0.0835   \\
		HSCNN-D	\cite{kaya2019towards}	 &3.61 M  	&5.22 T  & 40.55	&0.9836 &5.59    &0.0110  \\
        HIR-Net \cite{fu2018joint} &2.10 M     &2.94 T  & 39.80 &0.9861 &5.70    &0.0397 \\
        3D-CNN  \cite{koundinya20182d} &0.78 M &8.32 T  & 42.25	&0.9872 &5.24    &0.0093 \\
		FM-Net 	\cite{zhang2020pixel} &11.79 M &17.07 T	& 41.34 &0.9881	&6.09    &0.0101  \\
		AWAN \cite{li2020adaptive} &17.45 M    &24.63 T	& 43.35	&0.9919	&4.93	 &0.0089  \\
		Ours    &\textbf{0.22} M      &\textbf{0.76} T  & \textbf{43.97} &\textbf{0.9922} &\textbf{4.82}  &\textbf{0.0077} \\
		\hline
		\hline
	\end{tabular}}
\end{table*}

\begin{table*}
	\centering
	\caption{\label{tab:results2} Quantitative comparisons of different methods on the \textbf{CAVE} dataset. ``$\uparrow$ (resp. $\downarrow$)" indicates that the larger (resp. smaller), the better.  For \# Params and \# FLOPs, the smaller, the more compact and efficient. The best results are highlighted in bold.}
\resizebox{0.7\textwidth}{!}{
	\begin{tabular}{l|cc|cccc }
		\hline
		\hline
		\textbf{Methods} &  \# Params& \# FLOPs  & PSNR $\uparrow$ & ASSIM $\uparrow$ & SAM $\downarrow$ & RMSE $\downarrow$\\
		\hline
		\hline
		BI		   &--		&--	    &23.73  &0.8278   &33.81 &0.0877 \\
		HSCNN-D	\cite{kaya2019towards}	 &3.61 M	&0.95 T  &35.63  &0.9733   &9.63  &0.0194\\
        HIR-Net \cite{fu2018joint}       &2.10 M	&0.53 T  &33.97  &0.9456   &9.40  &0.0263\\
        3D-CNN  \cite{koundinya20182d}   &0.78 M    &1.53 T  &35.98  &0.9739   &8.89  &0.0182\\
		FM-Net 	\cite{zhang2020pixel}    &11.47 M	&3.09 T  &36.84  &0.9644   &8.54  &0.0179\\
		AWAN \cite{li2020adaptive}       &17.45 M	&4.57 T  &38.41  &0.9904   &8.08  &0.0170\\
		Ours                             &\textbf{0.26} M	&\textbf{0.14} T  &\textbf{39.68}  &\textbf{0.9894}   &\textbf{6.60}  &\textbf{0.0138}\\
		\hline
		\hline
	\end{tabular}}
\end{table*}

\begin{table*}
	\centering
	\caption{ Quantitative comparisons of different methods 
	on the \textbf{NTIRE 2020} dataset. ``$\uparrow$ (resp. $\downarrow$)" indicates that the larger (resp. smaller), the better.  For \# Params and \# FLOPs, the smaller, the more compact and efficient. The best results are highlighted in bold.}
	\label{tab:result3}
\resizebox{0.7\textwidth}{!}{
	\begin{tabular}{l|cc| cccc}
		\hline
		\hline
		\textbf{Methods} & \# Params & \# FLOPs   &PSNR $\uparrow$& ASSIM $\uparrow$& SAM $\downarrow$& RMSE $\downarrow$\\
		\hline
		\hline
		BI		                      &--	  &--	    &30.85 &0.9075 &8.48 &0.0394\\
		HSCNN-D	\cite{kaya2019towards}&3.61 M  &0.890 T   &41.42 &0.9946 &3.17 &0.0120 \\
        HIR-Net \cite{fu2018joint}   &2.01 M  &0.532 T   &35.26 &0.9862 &4.27 &0.0190\\
        3D-CNN	\cite{koundinya20182d}&0.78 M  &1.440 T   &40.81 &0.9938 &3.12 &0.0124 \\
		FM-Net	\cite{zhang2020pixel} &11.47 M &2.955 T   &42.36 &0.9950 &3.10 &0.0118\\
		AWAN	\cite{li2020adaptive} &17.45 M &4.300 T      &41.99 &0.9948 &3.22 &0.0112\\
		Ours	                      &\textbf{0.51} M  &\textbf{0.258} T &\textbf{43.39} &\textbf{0.9953} &\textbf{2.75} &\textbf{0.0101}\\
		\hline
		\hline
	\end{tabular}
	}
\end{table*}

\section{Experiments}
\label{sec:EXP}

\subsection{Experiment Settings and Implementation Details} 

We used 3 widely-used benchmark datasets 
i.e., HARVARD\footnote{http://vision.seas.harvard.edu/hyperspec/} \cite{chakrabarti2011statistics}, CAVE\footnote{http://www.cs.columbia.edu/CAVE/databases/} \cite{yasuma2010generalized}, and NTIRE 2020\footnote{http://www.vision.ee.ethz.ch/ntire20/} \cite{arad2020ntire}:
\begin{itemize}
    \item The CAVE dataset consists of 32 HS images of spatial dimensions 512 $\times$ 512 and spectral bands 31  captured by a generalized assorted pixel camera with an interval wavelength of 10nm in the range of 400-700nm. We randomly selected 20 HS images as the training set and the remaining 12 as the testing set. Following \cite{wei2020boosting}, \cite{zhang2020pixel}, we generated input RGB images using the camera spectral response function of Nikon D700.

    \item  The HARVARD dataset contains 50 indoor and outdoor HS images of spatial dimensions $1024\times 1392$ and spectral bands 31 covering 420-720 nm, which were captured under the daylight illumination. We utilized the first 30 HS images as the training set and the remaining 20 ones as the testing set. 
    Following \cite{wei2020boosting}, \cite{zhang2020pixel}, we generated input RGB images using the camera spectral response function of Nikon D700.
    
    \item The NTIRE 2020 dataset contains 450 HS/RGB image pairs from training, 10 pairs for validation, and 20 pairs for test. The HS images have 31 spectral bands covering 400-700nm. As the ground-truth images of the test set are unavailable, we adopted the validation set for evaluation. 
\end{itemize}

We adopted the ADAM \cite{kingma2014adam} optimizer with the exponential decay rates $\beta_1 = 0.9$ and $\beta_2 = 0.999$ for the first and second moment estimates, respectively. We initialized the learning rate of our AGD-Net as $1e-3$ and employed the cosine annealing decay strategy to gradually decrease it to $1e-5$. We empirically set $h_1$, $w_1$, $\delta_l$ and $\delta_h$ in Algorithm \ref{alg:forward} to 48, 48, 1e-3 and 1, respectively. During training, we fixed the same number of training epochs to 500 for all experiments. We implemented the model with PyTorch, and set the batch size to 8 for CAVE and HARVARD and 6 for NTIRE 2020.  



For a comprehensive quantitative evaluation, we adopted 4 commonly-used quantitative metrics, i.e., Peak Signal-to-Noise Ratio (PSNR), Average Structural Similarity Index (ASSIM) \cite{wang2002universal}, Spectral Angle Mapper (SAM) \cite{yuhas1992discrimination}, and Root Mean Squared Error (RMSE), which are respectively defined as: 

	\begin{equation}
\text{PSNR}(\mathbf{X},\widehat{\mathbf{X}}) = -\frac{10}{s} \sum_{c=1}^{s} \log(\text{MSE}(\mathbf{x}_c,\widehat{\mathbf{x}}_c)),
	\end{equation}
where $\widehat{\mathbf{x}}_c\in\mathbb{R}^{1\times hw}$ and $\mathbf{x}_c\in\mathbb{R}^{1\times hw}$  are the $c$-th ($1\leq c\leq s$) spectral bands of $\widehat{\mathbf{X}}$ and $\mathbf{X}$, respectively, $\text{MSE}(\cdot,\cdot)$ computes the mean squared error between the inputs. 

	\begin{equation}
	\text{ASSIM}(\mathbf{X},\widehat{\mathbf{X}}) = \frac{1}{s} \sum_{c=1}^{s}\text{SSIM}(\mathbf{x}_c,\widehat{\mathbf{x}}_c),
	\end{equation}
	where $\text{SSIM}(\cdot,\cdot)$ \cite{wang2004image} computes the SSIM value of a typical spectral band.

	\begin{equation}
	\text{SAM}(\mathbf{X},\widehat{\mathbf{X}}) = \frac{1}{hw} \sum_{j=1}^{hw}\arccos{\left(\frac{<\widehat{\mathbf{x}}^j, \mathbf{x}^j>}{\|\widehat{\mathbf{x}}^j\|_2\|\mathbf{x}^j\|_2}\right)},
	\end{equation}
	where $\widehat{\mathbf{x}}^j\in \mathbb{R}^{s\times 1}$ and $\mathbf{x}^j\in \mathbb{R}^{s\times 1}$ are the spectral signatures of the $j$-th ($1\leq j\leq hw$) pixels of $\widehat{\mathbf{X}}$ and $\mathbf{X}$, respectively,  $\|\cdot\|_2$ is $\ell_2$ norm of a vector, and $<\cdot, \cdot>$ calculates the inner product of two vectors.  
	\begin{equation}
	\text{RMSE}(\mathbf{X},\widehat{\mathbf{X}}) =\frac{1}{s} 
	\sum_{i=1}^{s}\sqrt{\frac{1}{hw} \sum_{j=1}^{hw} (x_{i,j} -\widehat{x}_{i,j})^2},
	\end{equation}
	where 
	$ \widehat{x}_{i,j}$ and $x_{i,j}$ are the $(i, j)$-th elements of  $\widehat{\mathbf{X}}$ and $\mathbf{X}$, respectively,


In addition, we also added up the number
of neural network parameters (\# Param) and the number
of floating number operations per-inference (\# FLOPs) of DNN-based methods to compare their efficiency.

\begin{figure*}
	\centering
	\includegraphics[width=0.85\textwidth]{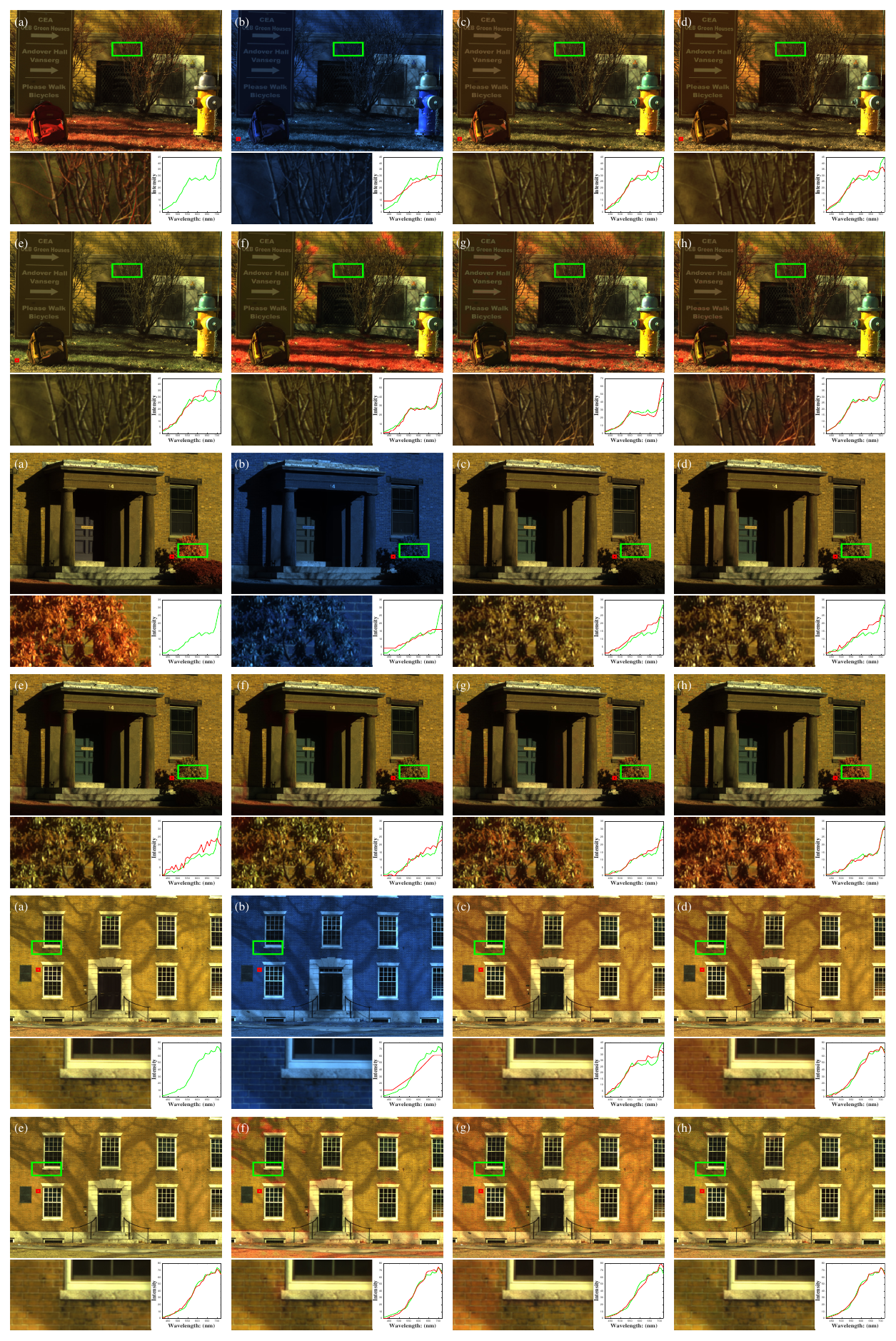}
	\caption{\label{fig:visual} Visual comparison of 3 HS images from the HARVARD dataset reconstructed by 7 different methods and the ground-truth images. To visualize HS images, 
	we extracted the $30th$, $20th$, and $10th$ spectral bands from an HS image as the red, green, and blue channels, respectively, to generate a pseudo-color image. (a) Ground-truth HS images, (b) Bicubic interpolation, (c) 3D-CNN, (d) HSCNN, (e) HIR-Net, (f) FM-Net, (g) AWAN, (h) Ours. For each subfigure, the bottom-left 
	is the zoomed-in patch indicated by the green frame in the pseudo-color image, and the bottom-right is the spectral curves of a typical pixel (red line) and its ground-truth (green line) 
	with the position of the selected pixel marked by the red rectangle in the pseudo-color image.}
\end{figure*}
\begin{figure*}
	\centering
	\includegraphics[width=0.85\textwidth,height=1.2\textwidth]{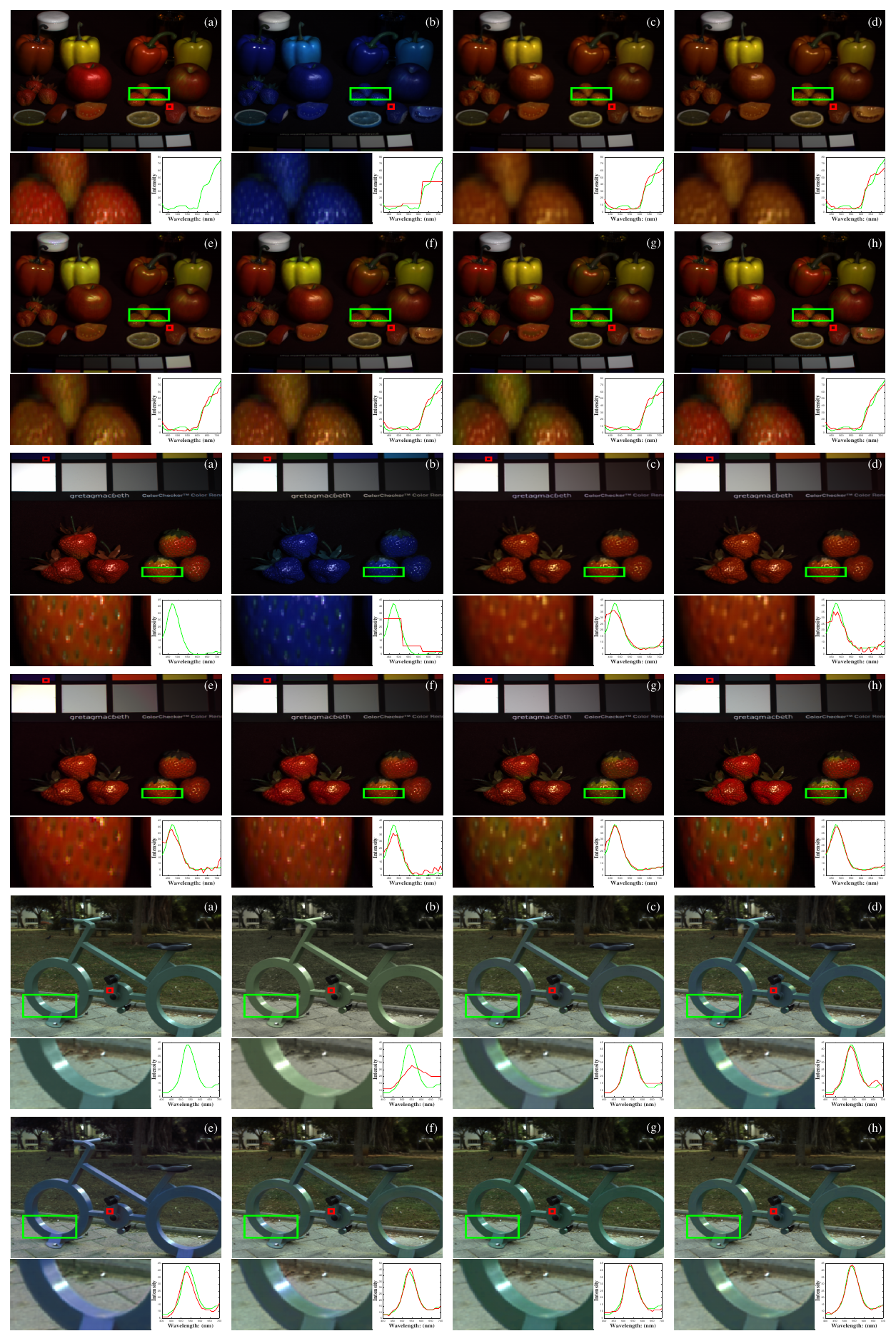}
	\caption{\label{fig:NTIRE_CAVE}  Visual comparison of 3 HS images reconstructed by 7 different methods and the ground-truth HS images. To visualize HS images, 
	we extracted the $30th$, $20th$, and $10th$ spectral bands from an HS image as the red, green, and blue channels, respectively, to generate a pseudo-color image. (a) Ground-truth HS images, (b) Bicubic interpolation, (c) 3D-CNN, (d) HSCNN, (e) HIR-Net, (f) FM-Net, (g) AWAN, (h) Ours. For each subfigure, the  bottom-left is 
	 the zoomed-in patch indicated by the green frame in the pseudo-color image, the bottom-right is the spectral curves of a typical pixel (red line) and its ground-truth (green line) 
	 with the position of the selected pixel marked by the red rectangle in the pseudo-color image. The top 2 and bottom 1 testing images from the CAVE and NTIRE 2020 datasets, respectively.}
\end{figure*}

\begin{table*}
\caption{\label{tab:ablation2} Results of ablation studies on the NTIRE 2020 dataset. 
$\times$ indicates 
that the corresponding component was removed when training AGD-Net. 
The bottom row corresponds to the complete AGD-Net. 
$1^{st}$ row: we utilized a linear convolutional layer of kernel size 1$\times$1 to replace the learned initialization module to learn the mapping from $\mathbf{Y}$ to $\widehat{\mathbf{X}}^{(0)}$; 
$2^{nd}$ row: we removed all sub-modules $\mathcal{D}_{(k)}(\cdot)$; 
$3^{rd}$ row: we learned the parameters of the convolutional layers $f_c\left(\cdot, \bm{\theta}_c^{(k)}\right)$ involved in different steps independently; 
$4^{th}$ row: we removed all SZM-norm layers; 
$5^{th}$ row: we removed the loss term $\mathcal{L}_F$ during training; 
$6^{th}$ row: we removed the loss term $\mathcal{L}_R$ during training; $7^{th}$ row: the full model.
}
\centering
\resizebox{0.9\textwidth}{!}{
\begin{tabular}{cccccc|cccc}
	\hline
	\hline
	Initialization &Incremental gradient & Sharing of $\bm{\theta}_{c}^{(K)}$ & SZM-norm & $\mathcal{L}_F$ &  $\mathcal{L}_R$  & PSNR $\uparrow$ & ASSIM $\uparrow$& SAM $\downarrow$& RMSE $\downarrow$ \\
	\hline
   $\times$    & $\checkmark$   & $\checkmark$    & $\checkmark$	& $\checkmark$  & $\checkmark$ &42.85 &0.9950 & 3.01&0.0116\\ 
   $\checkmark$    &$\times$    & $\checkmark$    & $\checkmark$	& $\checkmark$  & $\checkmark$ &42.40 &0.9945 &3.29 & 0.0115\\ 
   $\checkmark$    &$\checkmark$& $\times$        & $\checkmark$ & $\checkmark$  & $\checkmark$ &42.90 &0.9953  &2.98  &0.0106\\  
   $\checkmark$    &$\checkmark$& $\checkmark$    & $\times$	    & $\checkmark$  & $\checkmark$ &42.93 & 0.9956 & 2.98 & 0.0111\\ 
   $\checkmark$    &$\checkmark$& $\checkmark$    & $\checkmark$ & $\times$      & $\checkmark$ &41.69 & 0.9952 & 3.18 & 0.0122\\  
   $\checkmark$    &$\checkmark$& $\checkmark$    & $\checkmark$ & $\checkmark$  & $\times$     &43.02 & 0.9956 & 2.97 & 0.0110\\ 
   $\checkmark$    &$\checkmark$& $\checkmark$    & $\checkmark$ & $\checkmark$  & $\checkmark$ &43.39 & 0.9953 & 2.75 & 0.0101\\ 
	\hline
	\hline
\end{tabular}
}
\end{table*}

\begin{figure*}
	\centering
	\subfigure[PSNR ($\uparrow$)]{\includegraphics[width = 0.46\linewidth,
		height = 0.25\linewidth, trim=30 0 60 0, clip]{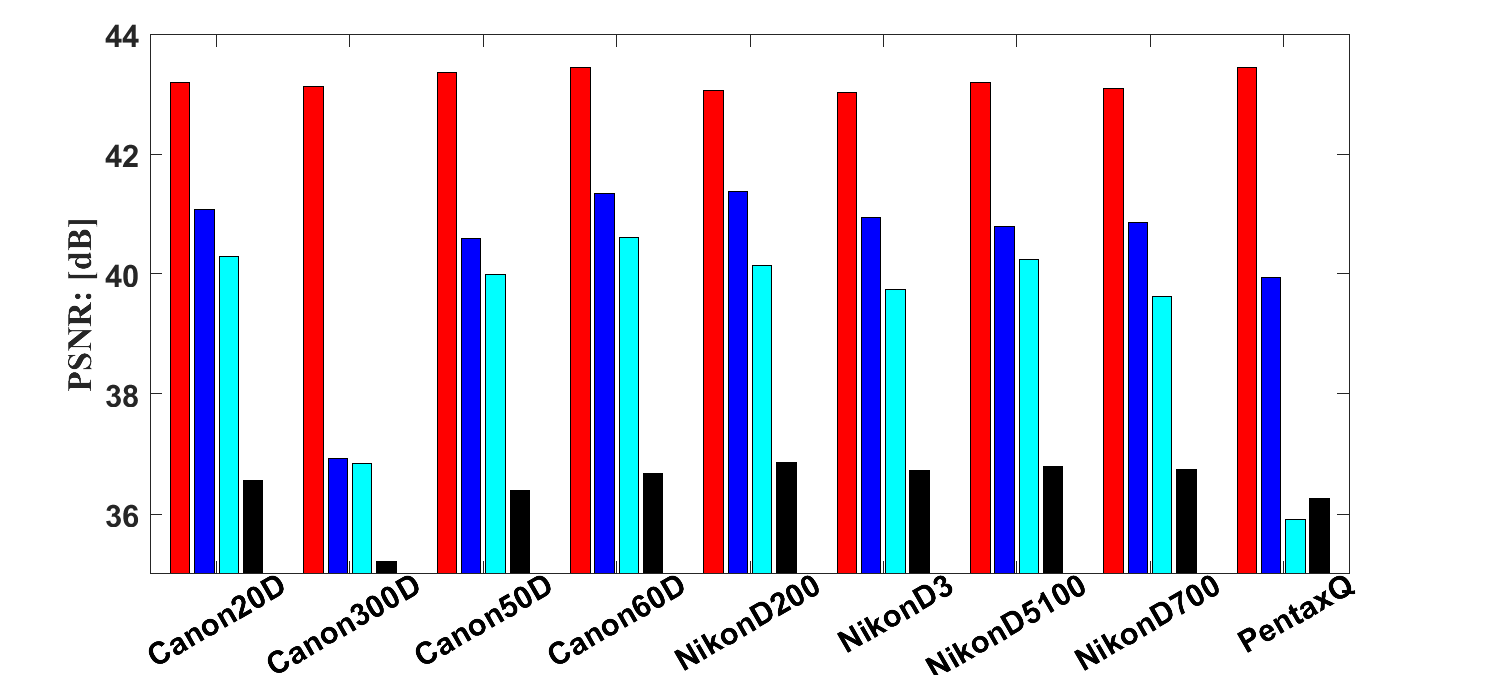}}
	\subfigure[ASSIM ($\uparrow$)]{\includegraphics[width = 0.46\linewidth,
		height = 0.25\linewidth,,trim=10 0 60 0, clip]{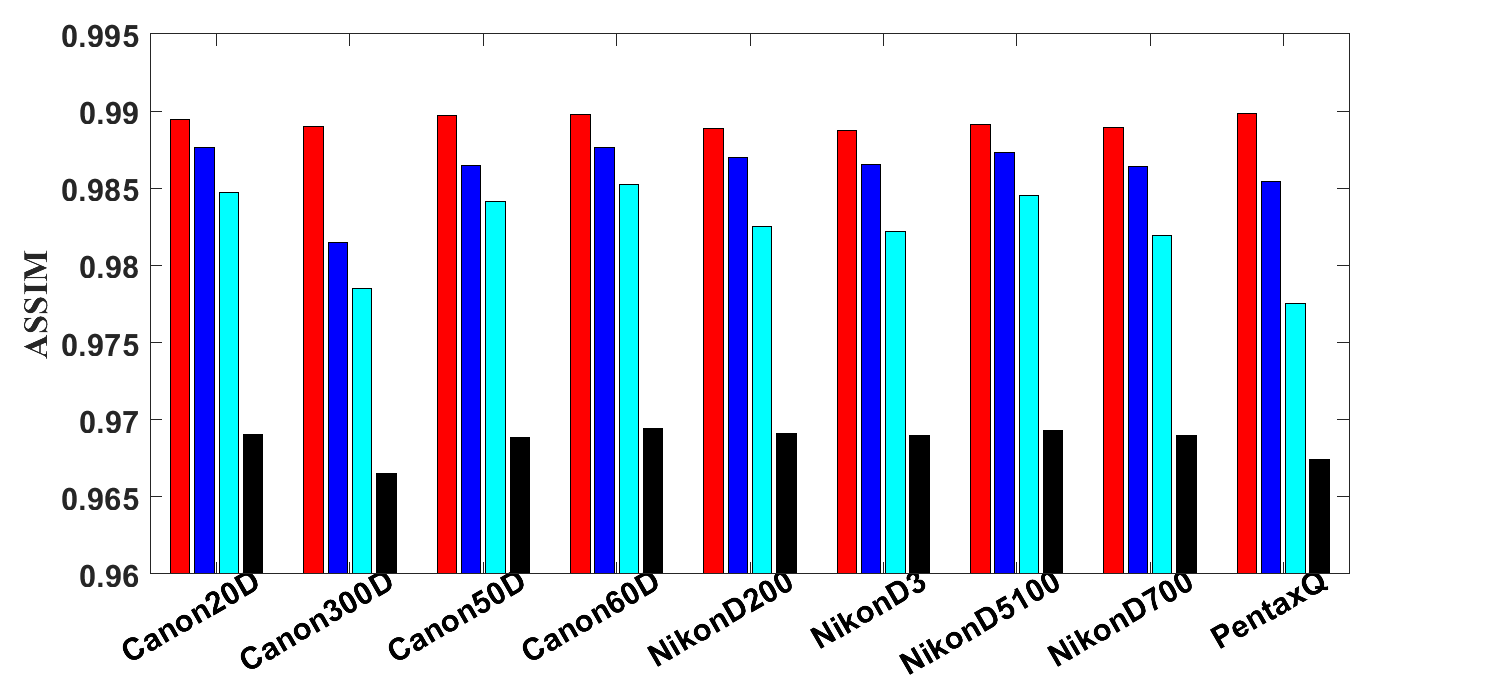}}
	\subfigure[SAM ($\downarrow$)]{\includegraphics[width = 0.46\linewidth,
		height = 0.25\linewidth,,trim=30 0 60 0, clip]{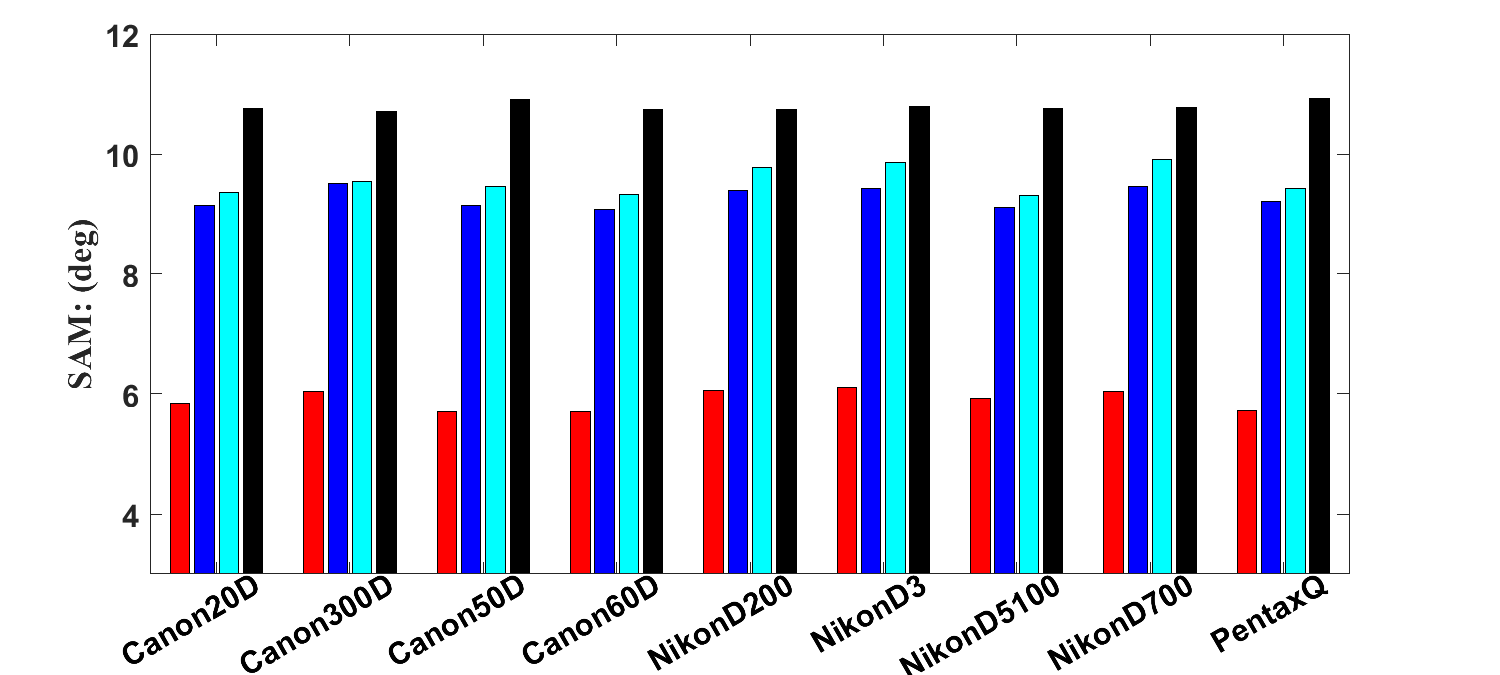}}
	\subfigure[RMSE ($\downarrow$)]{\includegraphics[width = 0.46\linewidth,
		height = 0.25\linewidth,,trim=30 0 60 0, clip]{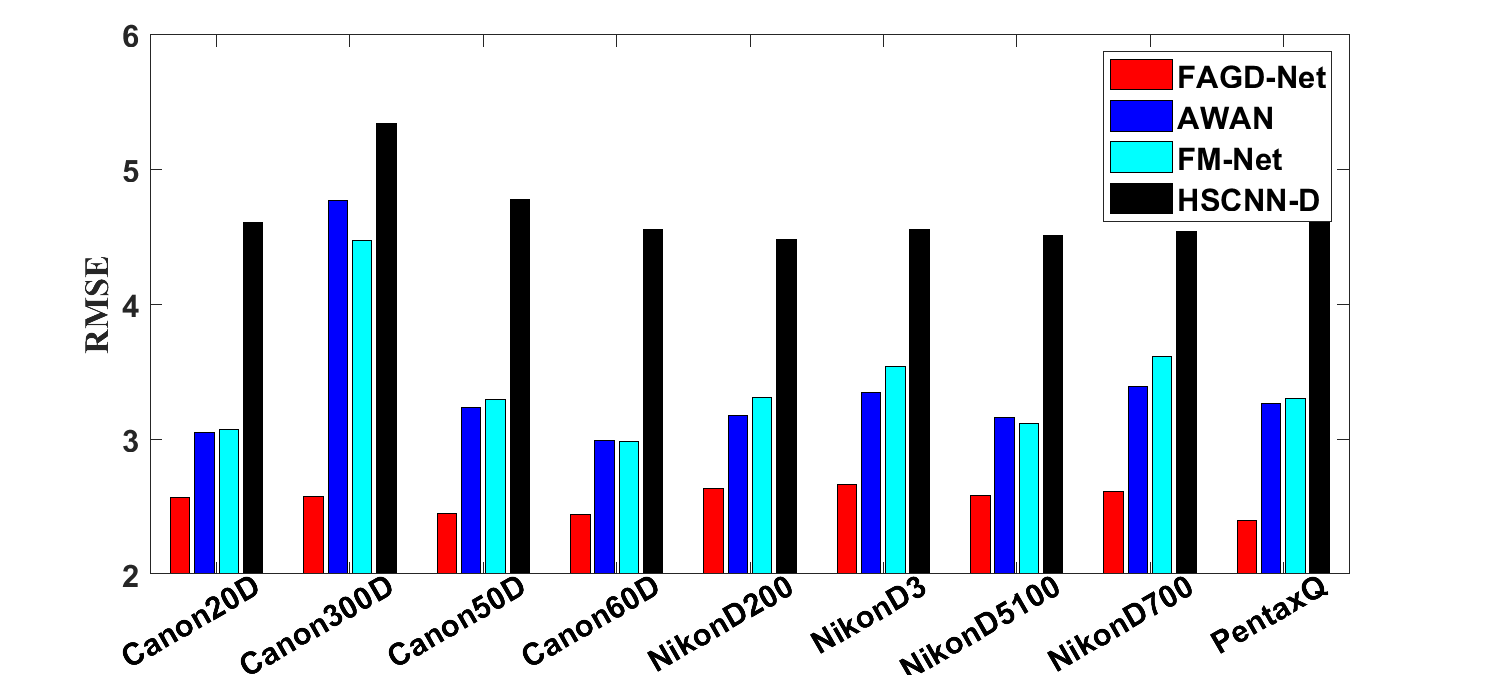}}
	\caption{\label{fig:extension} Quantitative comparison of different methods tested on RGB images obtained with 9 different SRFs listed in the horizontal axis to evaluate their flexibility. For each method, a single network was trained with RGB images obtained with 15 different SRFs. The four subfigures share the same legend shown in (d) 
	$\downarrow$ (resp. $\uparrow$) indicates the lower (resp. the higher), the better. 
	}
\end{figure*}
\subsection{Comparison with State-of-the-Art Methods}
\label{sec:exp}

We compared AGD-Net with 6 methods, including the Bicubic interpolation (BI) over the spectral dimension as a baseline and 5 most recent DNN-based methods, i.e., HSCNN-D \cite{shi2018hscnn+}, 3-D CNN \cite{koundinya20182d}, HIR-Net \cite{fu2018joint}, AWAN \cite{li2020adaptive}, and FM-Net \cite{zhang2020pixel}. 
Note that HSCNN-D and AWAN are the champion models of NTIRE 2018 \cite{8575291} and NTIRE 2020 \cite{koundinya20182d} challenge on spectral reconstruction from an RGB image, respectively.
For fair comparisons, we applied the same data pre-processing to all the methods, trained all the DNN-based methods
with the same training data by using the released codes 
with suggested parameters, and adopted the same protocol as \cite{dian2019learning, wang2019deep} to evaluate the experimental results of all the methods.

Tables \ref{tab:results1}, \ref{tab:results2} and \ref{tab:result3} list quantitative comparisons of different methods on the three benchmark datasets, where it can be seen that AGD-Net consistently surpasses all the compared methods in terms of all the four metrics, while consuming much fewer network parameters and FLOPs. 
Especially, AGD-Net improves PSNR by 1.27 dB (rep. 1.4 dB) and reduces SAM by $1.48^{\circ}$ (resp. $0.47^{\circ}$) on the CAVE (resp. NTIRE 2020) dataset, while saving more than 67$\times$ (resp. 32$\times$) parameters and 32$\times$ (resp. 16$\times$) FLOPs, as compared with the second-best method.

Figs. \ref{fig:visual} and \ref{fig:NTIRE_CAVE} visually compare different methods by showing their pseudo-color images and spectral curves, which still validate the significant superiority of our AGD-Net. 
Particularly, the compared methods cannot well handle regions either with high-frequency details (e.g., the branches in the $1^{st}$ image of Fig. \ref{fig:visual}, the flower patterns in the $2^{nd}$ image of Fig. \ref{fig:visual}, seeds of strawberries in the $2^{nd}$ image of Fig. \ref{fig:NTIRE_CAVE}) or smooth textures (e.g., the wall in the $3^{rd}$ image of Fig. \ref{fig:visual}, and the strawberries in $1^{st}$ image of Fig. \ref{fig:NTIRE_CAVE} ). By contrast, our AGD-Net produces much better results in these regions. Besides, the spectral curves by our method are closer to the ground-truth ones, e.g., 
the range of 600-720nm in the $2^{nd}$ image of Fig. \ref{fig:visual} , and the range of 500-720nm in $1^{st}$ image  of Fig. \ref{fig:NTIRE_CAVE}. 
Such advantages of AGD-Net are credited to that AGD-Net, built on an explicit observation model, is able to easily distinguish the high-frequency and low-frequency regions, and reconstruct them separately according to the projection errors. 

\subsection{Ablation Study}
We conducted extensive ablation studies to have a comprehensive understanding of AGD-Net.  

First, we experimentally validated the effectiveness of the initialization module, the learning of the incremental gradient, the manner of sharing projection coefficients $\theta_c$, the SZM-norm operation, and the loss function. 
As listed in Table \ref{tab:ablation2}, we can see that compared with the complete model, the reconstruction quality decreases after removing any one of these modules/operations, convincingly validating their effectiveness. Particularly, as listed in the $2^{nd}$ row, the PSNR drops about 1 dB without learning the incremental gradient, which demonstrates the rationality of our formulation of the amended gradient descent.
In addition, we observe that the self-supervised loss $\mathcal{L}_F$ makes significant contributions to the reconstruction process. The reason it that such a loss not only regularizes output HS images 
but also forces the network to regress the SRF for the correct calculation of the error maps in each module.

\begin{figure}
	\centering
	\subfigure[]{\includegraphics[width=0.8\linewidth]{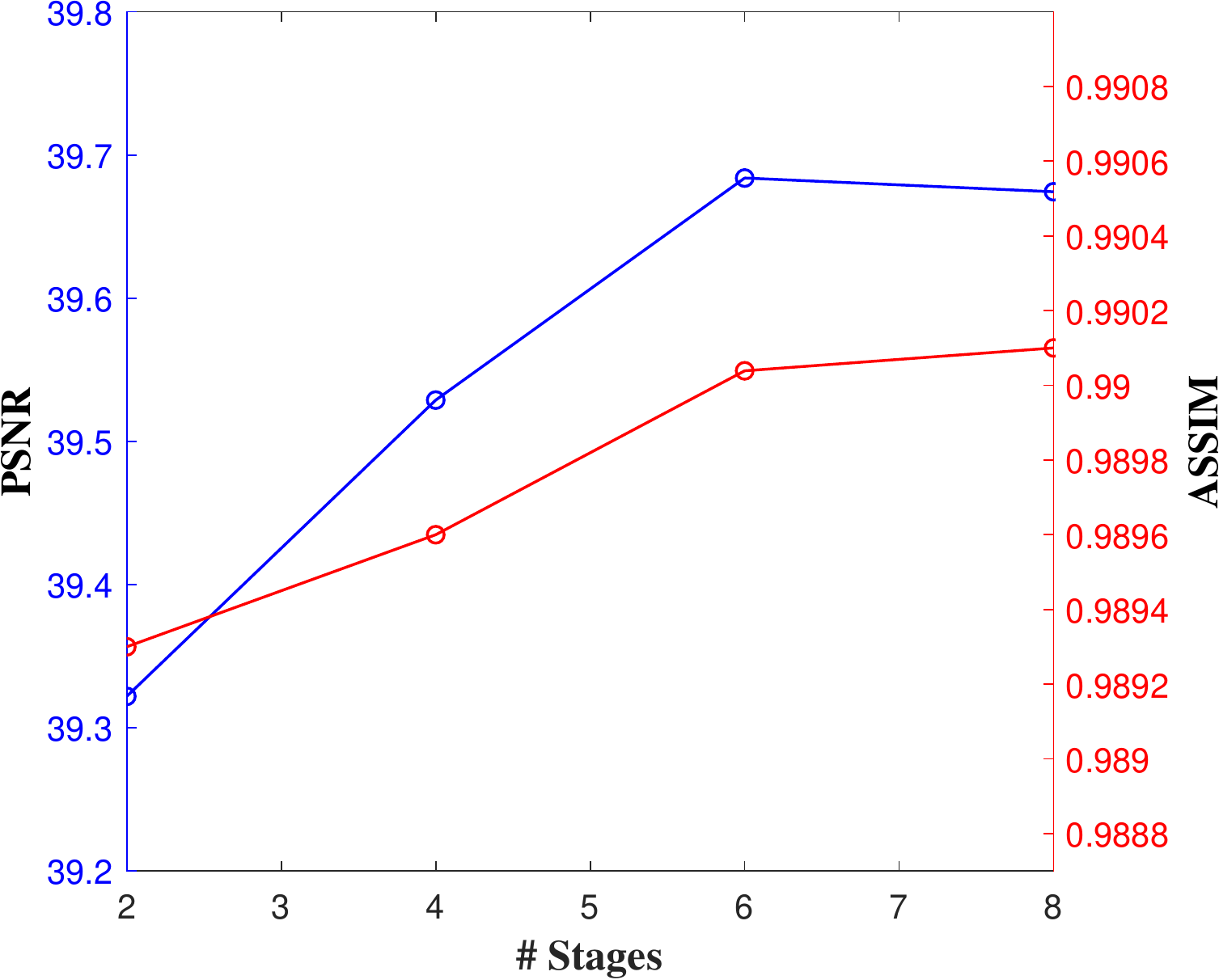}}
	\subfigure[]{\includegraphics[width=0.8\linewidth]{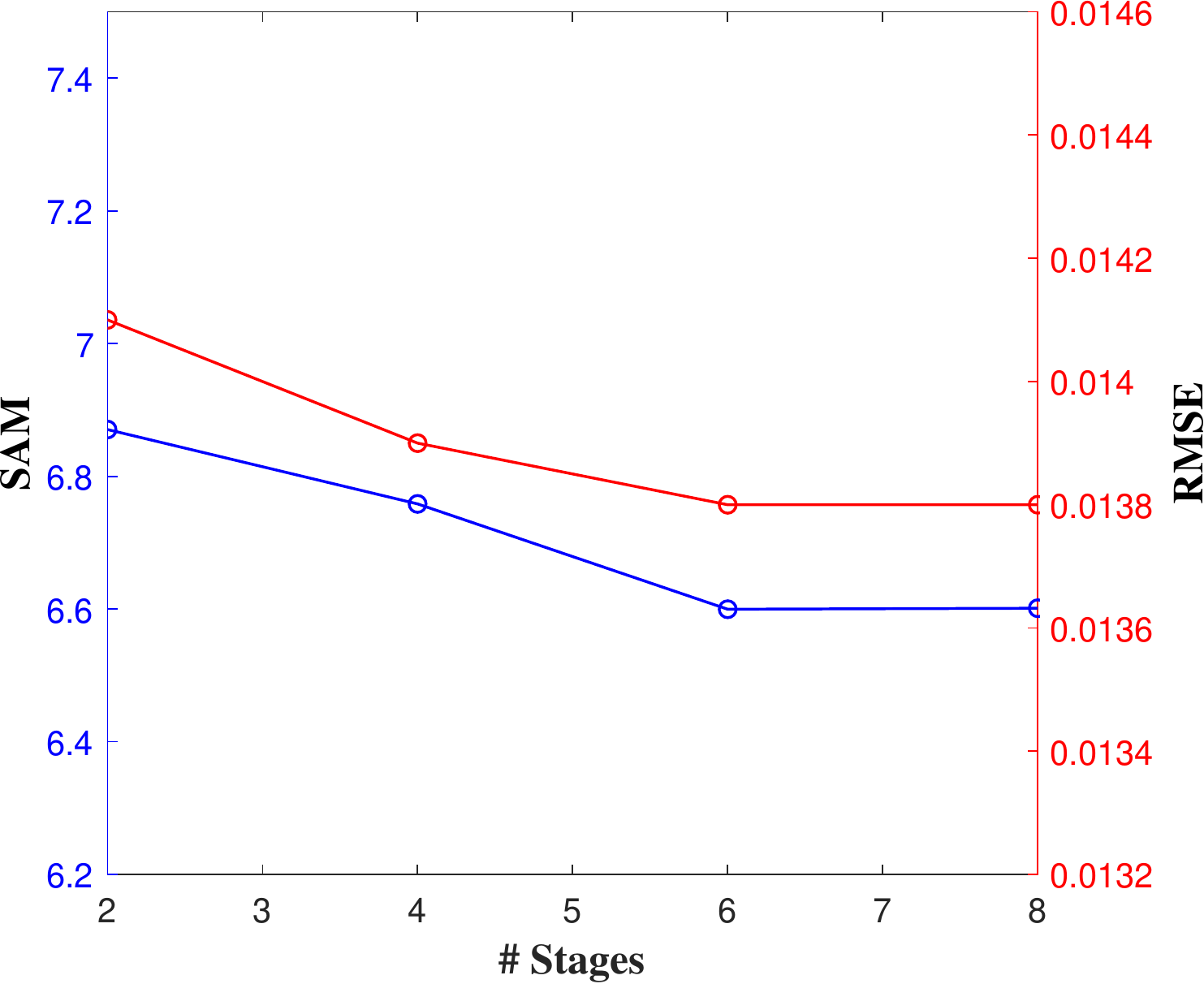}}
	\caption{\label{fig:ablation_stage} Investigation on the performance of AGD-Net with different number of stages on the CAVE dataset. 
	}
\end{figure}

We also investigated how the number of stages affects the reconstruction performance. 
Note that the initialization module is also considered as one stage. 
As shown in the Fig. \ref{fig:ablation_stage}, we can see that the performance of AGD-Net in terms of all the four metrics gradually improves with the number of stages increasing and gets saturated at 6 stages.
Thus, in all experiments, we set $K$ to 5, 6, and 12 stages for HARVARD, CAVE, and NTIRE datasets, respectively.


\subsection{Evaluation of the FAGD-Net}
\label{sec:expextension}



We used spectral response functions (SRFs) of 15 different cameras \cite{jiang2013space} to construct the training set, i.e., Canon1DMarkIII, Canon5DMarkII, NikonD300s, NikonD50, NokiaN900, Canon40D, Canon600D, NikonD3X, NikonD80, PhaseOne, Canon 500D, HasselbladH2, NikonD40, NikonD90,  and PointGreyGrasshopper214S5C. We generated the testing RGB images using SRFs of cameras Canon20D, Canon50D, NikonD200, NikonD5100, PentaxQ, Canon300D, Canon60D, NikonD3, and NikonD700.
The first 30 HS images from the HARVARD dataset were used for training, and the remaining 20 ones for testing. As the compared methods cannot utilize an SRF in an explicit manner, we projected the 30 HS images with respect to the 15 training SRFs to generate 450 pairs of HS and RGB images to train them. Note only a single network was trained for each method. 

Fig. \ref{fig:extension} shows the quantitative comparison of different methods, where
it can be seen that our FAGD-Net consistently exceeds the other methods to a significant extent on all 9 SRFs, 
e.g., the improvement of PSNR achieves 5.5 dB, and the reduction of SAM achieves about 3$\degree$ on Canon300D,  
validating the strong flexibility or generalization ability of FAGD-Net  to different SRFs, which is credited to the interpretable network architecture.

\section{Conclusion}
\label{sec:CON}
We have presented AGD-Net,
a novel end-to-end learning
framework for the reconstruction of HS images from single RGB images. As a neural network built upon an explicit formulation of using the gradient descent algorithm, AGD-Net is interpretable and compact.  In addition to the blind reconstruction, i.e., SRFs are unknown, AGD-Net is also adapted to 
non-blind reconstruction by explicitly utilizing known SRFs,  distinguishing itself from the deep learning peers in flexibility: trained once a single
network of AGD-Net is able to well handle input RGB images obtained via different SRFs. 
We demonstrated the significant advantages of AGD-Net over state-of-the-art methods by conducting extensive experiments as well as comprehensive ablation studies. That is, 
AGD-Net improves PSNR up to 5.5 dB and reduces SAM up to 3$\degree$ while saving up to 67$\times$ parameters and 32$\times$ FLOPs. 
We believe our new perspective will bring insights to other inverse problems in image processing,
such as image super-resolution, deblurring, and compressive sensing.

\bibliographystyle{IEEEtran}
\bibliography{egbib}
\end{document}